\documentclass[12pt]{article}
\include{psfig}

\def\half{{\textstyle{1\over2}}}
\def\quarter{{\textstyle{1\over4}}}
\def\sixth{{\textstyle{1\over6}}}
\def\CC{{\cal C}}
\def\CD{{\cal D}}
\def\CF{{\cal F}}
\def\CL{{\cal L}}
\def\CO{{\cal O}}
\def\CS{{\cal T}}

\def\CV{{\cal V}}

\def\ixi#1{\overline{(I\otimes\xi_{#1})} }
\def\sfno#1#2{\overline{(\gamma_{#1}\otimes\xi_{#2})}}
\def\semitimes{\mathrel>\joinrel\mathrel\triangleleft}
\def\slash#1{\mbox{$\not \!\! #1$}}

\def\Tr{{\sf Tr}}

\def\bar{\overline}

\def\tilde{\widetilde}

\def\spose#1{\hbox to 0pt{#1\hss}}
\def\ltapprox{\mathrel{\spose{\lower 3pt\hbox{$\mathchar"218$}}
 \raise 2.0pt\hbox{$\mathchar"13C$}}}
\def\gtapprox{\mathrel{\spose{\lower 3pt\hbox{$\mathchar"218$}}
 \raise 2.0pt\hbox{$\mathchar"13E$}}}
\def\inapprox{\mathrel{\spose{\lower 3pt\hbox{$\mathchar"218$}}
 \raise 2.0pt\hbox{$\mathchar"232$}}}

\begin{document}
\begin{titlepage}

\begin{flushright}
\vspace{-0.3truein}
UW/PT-98-13\\
UTCCP-P-57
\end{flushright}


\begin{center}
\LARGE{\bf Partial Flavor Symmetry Restoration \\
		for Chiral Staggered Fermions}
\end{center}
\vspace{0.1in}
\begin{center}
 {%
  \begin{tabular}[t]{c}
        {\large Weonjong Lee\footnotemark}\\[1em]
        \em Los Alamos National Laboratory, Mail Stop B-285,
        Los Alamos, NM 87545, USA \\[3em]
        {\large Stephen R. Sharpe \footnotemark}\\[1em]
        \em Physics Department, Box 351560, University of Washington,
        Seattle, WA 98195-1560, USA \\[.5em]
	and \\[.5em]
	\em Center for Computational Physics, University
                 of Tsukuba, Tsukuba, Ibaraki, 305-8577, Japan
  \end{tabular}}
\end{center}

\vspace{.5cm}

\begin{small}
\centerline{ABSTRACT}
\medskip

We study the leading discretization errors for staggered fermions
by first constructing the continuum effective Lagrangian including
terms of $O(a^2)$, and then constructing the corresponding
effective chiral Lagrangian. 
The terms of $O(a^2)$ in the continuum effective Lagrangian
completely break the $SU(4)$ flavor symmetry down to the discrete
subgroup respected by the lattice theory.
We find, however, that the $O(a^2)$ terms in the potential
of the chiral Lagrangian maintain an $SO(4)$ subgroup of $SU(4)$. 
It follows that the leading
discretization errors in the pion masses are $SO(4)$ symmetric,
implying three degeneracies within the seven lattice
irreducible representations. 
These predictions hold also for perturbatively improved versions of
the action.
These degeneracies are observed,
to a surprising degree of accuracy, in existing data.
We argue that the $SO(4)$ symmetry does not extend to the masses
and interactions of other hadrons (vector mesons, baryons, etc),
nor to higher order in $a^2$.
We show how it is possible that, for physical quark masses of $O(a^2)$,
the new $SO(4)$ symmetry can be spontaneously broken, leading to
a staggered analogue of the Aoki-phase of Wilson fermions.
This does not, however, appear to happen for presently studied versions of
the staggered action.

\end{small}
\bigskip
\centerline{(submitted to Physical Review D)}

\footnotetext[1]{Email:    wlee@lanl.gov}
\footnotetext{Email:    sharpe@phys.washington.edu}
\vfill
\end{titlepage}

\section{Introduction}\label{sec:intro}

Staggered fermions~\cite{susskind} are
one of the standard discretizations used in lattice simulations. 
Their major advantage is that they retain a remnant
of the continuum chiral symmetry group at non-zero lattice spacing.
Their major drawback is that the continuum flavor symmetry
($SU(4)$ for a single staggered fermion)
is almost completely broken on the lattice---only a
discrete subgroup remains.
Although only an effect of $O(a^2)$, flavor breaking
turns out to be numerically significant at typical lattice spacings.

In this paper we show that, in the Goldstone boson sector,
flavor symmetry is partially restored in the chiral limit.
The discrete lattice flavor group enlarges to an $SO(4)$ subgroup of
the continuum $SU(4)$. This implies relations between
multi-pion interactions at very low momenta.
The simplest of these concerns the pion masses: seven 
lattice irreducible representations (irreps) collapse into four $SO(4)$
irreps in the chiral limit, implying three degeneracies.
This result applies to both the original form of staggered fermions,
and to various improvements that have been recently
suggested~\cite{toussaint,lagae,orginos,lepageimp}.
It turns out that these degeneracies are present both in data from
a few years ago~\cite{ishizuka}, and from recent simulations~\cite{orginos}.

Our derivation of these degeneracies applies only to the leading order
discretization errors. We find that they are violated by terms of
$O(a^4)$. We have also studied the extension of our result to the
masses and interactions of hadrons other than the pions---we
find that the partial flavor symmetry restoration does not
extend to the masses and interactions of the vector mesons, and
we argue that this is also true for any
particle which remains massive in the chiral and continuum limits. 
In other words, for such particles all lattice irreps should be split
by terms of $O(a^2)$, even in the chiral limit.
Present numerical results for vector mesons and baryons are either
of insufficient accuracy or do not consider enough irreps
to test this prediction.

In the bulk of this article we explain how the enlarged flavor
symmetry is derived. The method is modeled on that used to
study flavor and parity breaking with Wilson fermions~\cite{ss}.
While straightforward in principle,
the application of the method to staggered fermions is considerably
more complicated.
One begins by determining all the 
lattice operators which are required for $O(a^2)$ improvement.
This has largely been done by Luo~\cite{luo2}, 
but he missed some operators, 
and we complete the task in App.~\ref{app:ops}.
The second step is to write down
the effective continuum action which describes the 
lattice theory including errors proportional to $a^2$,
and determine the symmetries of its various components.
This is done in Sec.~\ref{sec:effaction}.
A side benefit of this analysis is seeing explicitly how the
lattice symmetry group emerges in continuum language.
One then writes down the effective chiral Lagrangian describing
the interactions of the pseudo-Goldstone bosons in
the effective continuum theory, again including the $a^2$ terms.
This is worked out in Sec.~\ref{sec:chiral}
(with details relegated to App.~\ref{app:matching}), 
and it is at this stage that the symmetry restoration emerges.
The implications of partial symmetry restoration
for pion masses are presented in Sec.~\ref{sec:pion},
and are compared to the results of numerical simulations.
In Sec.~\ref{sec:nonGoldstone} we explain
why the partial symmetry restoration does not extend to vector mesons.

It turns out that the enlarged flavor symmetry which emerges in the
chiral limit at $O(a^2)$ can also be spontaneously broken in that same limit.
Whether this happens depends on the sign of various unknown coefficients
in the chiral Lagrangian describing the lattice theory.
This phenomenon is analogous to the breaking of flavor in the Aoki-phase
for Wilson fermions~\cite{aoki,ss}.
The analogy is not precise, however, because the symmetry that is being broken 
here is not an exact symmetry of the lattice theory, but is instead
violated by terms of $O(a^4)$. Thus there are only pseudo-Goldstone bosons
in the broken phase.
The new phase, if it occurs, is very narrow---in terms of the bare lattice
quark masses it has a width of $O(a^3)$.
It appears that this phase is unlikely to occur for those versions
of staggered fermions presently in use, but it may occur
for other improved versions.
In Sec.~\ref{sec:phase} we give a brief discussion of the
properties of the new phase.

We close with some comments on the practical implications of our result.

A preliminary report on this work 
was presented in Ref.~\cite{leesharpelat98}. 
At that time we suggested that the enlarged symmetry would apply also
to hadrons other than the pions, a possibility we now think is incorrect.

\section{The Continuum Effective Action}\label{sec:effaction}

The first step in our analysis is to write down the effective 
continuum action describing the interactions of quarks and gluons with
momenta much below the cut-off, $p\ll \Lambda \sim 1/a$ \cite{symanzik}.
Integrating out modes near the cut-off introduces higher dimension operators
with coefficients suppressed by explicit powers of $a$.
The constraint on these operators is that they must have the same
symmetries as the underlying lattice action.
We wish to include terms suppressed by up to two powers of $a$,
and thus must determine
all allowed operators of dimension up to and including six.
This we do by writing down the allowed lattice operators
and then matching them onto continuum operators at tree level.
The result has the form
\begin{eqnarray}
S_{\rm eff} &=& S_4 + a^2 S_6  \\
&=& \int d^4x ({\CL}_4 + a^2 {\CL}_6) \,.
\end{eqnarray}
The leading term is 
\begin{equation}
\CL_4 = \frac12 {\sf Tr} F_{\mu\nu} F_{\mu\nu}
	+ \bar Q (\slash D + m) Q \,,
\end{equation}
with 
\begin{equation}
\slash D = \sum_\mu (\gamma_\mu \otimes 1) D_\mu \,,
\end{equation}
in a $(\hbox{\rm spin}\otimes\hbox{\rm flavor})$ notation
explained in more detail in App.~\ref{app:ops}.
The physical quark mass is proportional to the bare staggered
quark mass $m (\mu) a = Z_m(g^2,\ln (a\mu)) m_0(a)$.
There are no terms linear in $a$ because there are no operators
of dimension five consistent with all the lattice symmetries.
This is well known for the gauge theory, and has been shown in
Refs.~\cite{sslat93,luo1} for staggered fermions.

There are three types of contribution to $S_6$:
gluonic operators, fermion bilinears and four-fermion operators
\begin{equation}
S_6 = S_6^{\rm glue} + S_6^{\rm bilin} + S_6^{\rm FF} \,.
\end{equation}
The gluonic terms are~\cite{LW}
\begin{equation}
\CL_6^{\rm glue} \sim
{\sf Tr}\left(D_\mu F_{\mu\nu} D_\rho F_{\rho\nu}\right) + 
\sum_{\mu\nu\rho}
{\sf Tr}\left(D_\mu F_{\nu\rho} D_\mu F_{\nu\rho}\right) +
 \sum_{\mu\nu}
{\sf Tr}\left(D_\mu F_{\mu\nu} D_\mu F_{\mu\nu}\right) \,,
\label{eq:L6glue}
\end{equation}
where the symbol $\sim$ implies that each of the operators
appears with a coefficient which depends on $g^2$ and $\ln a$.
The first term in eq.~(\ref{eq:L6glue}) can be absorbed into
a redefinition of gluon field,
which leaves only two independent terms in $ \CL_6^{\rm glue} $.

The enumeration of the fermionic operators is described in 
App.~\ref{app:ops}, along with the notation used to refer to them.
Most have been previously listed by Luo~\cite{luo2}.
There are 8 fermion bilinears
\begin{eqnarray}
\CL_6^{\rm bilin} &\sim&
\bar Q (\slash D)^3 Q +
\sum_\mu \bar Q \left(D_\mu^2 \slash D + \slash D D_\mu^2\right) Q
\nonumber \\
&&\mbox{}+ 
\sum_\mu \bar Q D_\mu \slash D D_\mu Q +
\sum_\mu \bar Q (\gamma_\mu\otimes1) D_\mu^3 Q 
\nonumber \\
&&\mbox{}+
m\bar Q (\slash D)^2 Q + \sum_\mu m\bar Q D_\mu^2 Q
+ m^2 \bar Q \slash D Q + m^3 \bar Q Q \,,
\label{eq:L6bilin}
\end{eqnarray}
and 24 four-fermion operators. The latter we divide into two classes:
\begin{eqnarray}
\CL_6^{\rm FF(A)} &\sim&
\,[S\times A] + [S\times V] + [A\times S] + [V\times S] +\nonumber \\
&&\mbox{}+
\,[P\times V] + [P\times A] + [V\times P] + [A\times P] +\nonumber \\
&&\mbox{}+
\,[T\times V] + [T\times A] + [V\times T] + [A\times T] +\nonumber \\
&&\mbox{}+
\left\{[S\times S]-[P\times P]\right\} + 
\left\{[S\times P]-[P\times S]\right\} + \nonumber \\
&&\mbox{}+
\left\{[S\times T]-[P\times T]\right\} + 
\left\{[T\times S]-[T\times P]\right\} + \nonumber \\
&&\mbox{}+
\left\{[V\times V]-[A\times A]\right\} + 
\left\{[V\times A]-[A\times V]\right\}
\label{eq:L6FFA}
\end{eqnarray}
(with compound operators enclosed in curly braces), and
\begin{eqnarray}
\CL_6^{\rm FF(B)} &\sim&
\,[T_\mu\times V_\mu] + [T_\mu\times A_\mu] + [V_\mu\times T_\mu] +
\,[A_\mu\times T_\mu] +\nonumber \\
&&\mbox{}+
\left\{([V_\mu\times V_\mu]-[A_\mu\times A_\mu]) -\quarter
([V\times V]-[A\times A])\right\} \nonumber \\
&&\mbox{}+
\left\{([V_\mu\times A_\mu]-[A_\mu\times V_\mu]) -\quarter 
([V\times A]-[A\times V])\right\} \,.
\label{eq:L6FFB}
\end{eqnarray}
The essential point of the notation is that the first letter indicates
the rotation property of the bilinears within the operator, while the
second indicates the flavor. Thus in
$[S\times T]$ both bilinears are scalars and have
``tensor'' flavor, i.e. a flavor matrix of the form 
$\xi_\mu\xi_\nu\equiv\xi_{\mu\nu}$\,.

Several of these operators are redundant, 
i.e. can be absorbed into the others by a field transformation. 
Although we could use this to reduce the number of operators,
we choose not to do so, since it is little extra work to keep
track of all the operators. In this way we do not have to worry
about the subtleties of determining redundant operators with staggered 
fermions~\cite{lepageimp}.

As noted in the introduction, a variety of improvements to the 
staggered fermion action are being tested in numerical simulations. 
All of these use perturbation theory,
and thus reduce, but do not remove, the errors proportional to $a^2$.
This implies that the form of the continuum effective action 
remains the same as for unimproved staggered fermions,
although the size of the numerical coefficients multiplying
the $a^2$ terms is presumably reduced.

\subsection{Symmetries of the effective action}

For the subsequent analysis, it is crucial to understand
the symmetries that are maintained by
the various terms in the effective action. 

The leading order action $S_4$ 
is invariant under Euclidean translations, rotations and reflections,
and also under charge conjugation $\CC$, fermion number $U(1)_V$, 
and the flavor group $SU(4)$. 
In the massless limit the flavor symmetry enlarges to the chiral
symmetry $SU(4)_L\times SU(4)_R$.
In general, we expect $S_6$ to break this symmetry down
to that of the underlying lattice theory,
because we have included in $S_6$ all operators consistent with the
lattice symmetries. While it is possible that some symmetries will
not be broken until higher order in $a$, 
this turns out not to be the case here.
The only exception is continuous 
translation invariance, which is a symmetry of $S_{\rm eff}$ but
is broken by the lattice to a discrete subgroup. This symmetry is
effectively restored because we are considering only low-momentum modes.

We first note that fermion number, charge conjugation and spatial
inversion (parity) are not broken by $S_6$. 
This is obvious for fermion number.
Charge conjugation does flip the sign of vector and tensor bilinears,
but this sign cancels since these bilinears always appear in pairs.
Parity also flips the sign of some bilinears, but again the 
overall sign cancels.

\begin{table}
\renewcommand{\arraystretch}{1.4}
\begin{center}
\begin{tabular}{ll}
\hline
Term in action & [Flavor] $\times$ Rotation symmetry \\ \hline
$S_4$ ($m=0$) 		& $[SU(4)_L\times SU(4)_R] \times SO(4)$ \\
$S_4$ ($m\ne0$) 	& $[SU(4)] \times SO(4)$ \\ 
\hline
$S_6^{\rm glue}$	& $[SU(4)_L\times SU(4)_R] \times SW_4$ \\
$S_6^{\rm bilin}$ ($m=0$)& $[SU(4)_L\times SU(4)_R] \times SW_4$ \\
$S_6^{\rm bilin}$ ($m\ne0$)& $[SU(4)] \times SW_4$ \\
$S_6^{\rm FF(A)}$	& 
$[U(1)_A\times\Gamma_4 \semitimes SO(4)] \times SO(4)$ \\
$S_6^{\rm FF(B)}$	& 
$U(1)_A\times (\Gamma_4 \semitimes SW_{4,\rm diag})$ \\
\hline
$S_6 (m=0)$		&
$U(1)_A\times (\Gamma_4 \semitimes SW_{4,\rm diag})$ \\
$S_6 (m\ne 0)$		&
$\Gamma_4 \semitimes SW_{4,\rm diag}$ \\
\hline
\end{tabular}
\caption{The flavor (including chiral) 
and rotation symmetries respected
by various terms in the effective action. 
In the last three lines flavor and rotation symmetries are intertwined.
See text for detailed explanation.
}
\label{tab:symm}
\end{center}
\end{table}

This leaves us to consider the breaking of flavor and rotation symmetry.
In Table~\ref{tab:symm} we display the symmetries
respected by each of the terms in the effective action.
The notation is as follows: $SW_4$ is the hypercubic subgroup of the Euclidean
rotation group $SO(4)$;
$\Gamma_4$ is the Clifford group with four generators which square to the
identity; $\semitimes$ indicates a semi-direct product; and $U(1)_A$
is the axial symmetry 
\begin{equation}
Q \to \exp\left(i \theta_A (\gamma_5\otimes\xi_5) \right) Q \,,\qquad
\bar Q \to \bar Q \exp\left(i \theta_A (\gamma_5\otimes\xi_5) \right)
\,,
\end{equation}
which is the continuum equivalent of the lattice axial symmetry.
The meaning of $SW_{4,\rm diag}$ will be described below.

We now explain the results in the Table, 
commenting on their significance as we proceed.
The gluonic action $S_6^{\rm glue}$ does not affect the flavor symmetry,
but the third term in eq.~(\ref{eq:L6glue}) breaks the $SO(4)$ rotation
symmetry down to its hypercubic subgroup $SW_4$, 
because the index $\mu$ is repeated four times. 
The same holds true for $S_6^{\rm bilin}$:
all terms in eq.~(\ref{eq:L6bilin}) are flavor singlets, but the
fourth term breaks $SO(4)\to SW_4$.

Flavor symmetry is broken only by the four-fermion terms.
This occurs in two stages. All terms in $S_6^{FF(A)}$ have their
Lorentz and flavor indices contracted separately, e.g.
\begin{equation}
[T\times V] \equiv
\sum_{\mu<\nu} \sum_\rho
\bar Q(y) (\gamma_{\mu\nu}\otimes\xi_{\rho}) Q(y)
\,\bar Q(y) (\gamma_{\nu\mu}\otimes\xi_{\rho}) Q(y)
\,,
\end{equation}
and are thus Lorentz singlets.
The appearance of flavor matrices implies that they break 
the $SU(4)$ symmetry, but the contraction of flavor indices between
bilinears implies that an $SO(4)$ subgroup survives unbroken.
This is the subgroup in which the {\bf 4} of $SU(4)$ transforms in
the spinor rep of $SO(4)$:
\begin{equation}
Q \longrightarrow (1 \otimes \Lambda_{1/2}) Q \,,\quad
\Lambda_{1/2} = \exp(\omega_{\mu\nu}\xi_{\mu\nu}) \,,
\end{equation}
with $\omega_{\mu\nu}$ an antisymmetric matrix of real parameters.
The appearance of operators containing bilinears of all possible flavors
in eq.~(\ref{eq:L6FFA}) implies that no other continuous flavor symmetry
remains. One can also show that no axial symmetries survive aside from
the $U(1)_A$ described above.

There is, however, an additional discrete flavor symmetry.
All the four-fermion operators (of both type A and B)
are invariant under the ``flavor reflections'' 
generated by the transformations\footnote{%
In Ref.~\cite{leesharpelat98} we used a different
basis for the generators, namely $(i\xi_{5\mu})$. 
Although the two choices of basis are equivalent, that we use here
allows a more direct connection to the lattice symmetries.}
\begin{equation}
Q \to (1\otimes \xi_\mu) Q \,,\quad
\bar Q \to \bar Q (1 \otimes \xi_\mu) \,.
\label{eq:gens}
\end{equation}
These transformations may change the sign of a bilinear,
\begin{equation}
\bar Q (\gamma_S\otimes \xi_F) Q \to
\bar Q (\gamma_S\otimes \xi_{\mu F \mu}) Q
= (-)^{\tilde F_\mu} \bar Q (\gamma_S\otimes \xi_F) Q
\label{eq:contdiscflav}
\end{equation}
(see App.~\ref{app:ops} for notation),
but this sign cancels in the four-fermion operators since all
bilinears appear in flavor diagonal pairs.
The generators in (\ref{eq:gens})
anticommute and square to $+1$, and so generate the
Clifford group $\Gamma_{4}$.
They lie within the flavor group $SU(4)$, but not within
the $SO(4)$ subgroup (which is composed of matrices containing an even
number of $\xi_\mu$).
They transform as a vector under the flavor $SO(4)$, and thus are
combined with this group in a semi-direct product as shown in the Table.

The second stage of symmetry breaking is due to the six operators
in $S_6^{FF(B)}$, in which the flavor and Lorentz indices are correlated.
Two examples appear in App.~\ref{app:ops}
[eqs.~(\ref{eq:VVdef}) and (\ref{eq:VTdef})]; here we give another
(this time in continuum notation)
\begin{eqnarray}
[T_\mu\times A_\mu] &\equiv& \sum_{\mu<\nu}
\bar Q(y) \sfno{\mu\nu}{\mu5} Q(y)
\,\bar Q(y) \sfno{\nu\mu}{5\mu} Q(y) \nonumber\\
&&\mbox{}\quad\quad -
\bar Q(y) \sfno{\mu\nu5}{\mu5} Q(y)
\,\bar Q(y) \sfno{5\nu\mu}{5\mu} Q(y)
\,.
\label{TAdef}
\end{eqnarray}
These operators are not invariant under either Lorentz or 
flavor $SO(4)$ rotations.\footnote{%
The extra subtractions in the last two compound operators in
eq.~(\ref{eq:L6FFB}) are included to remove a Lorentz singlet component.}
If, however, we do a simultaneous Lorentz and flavor {\em hypercubic}
rotation, then the correlation between the indices is preserved.
The restriction to hypercubic, rather than continuous, rotations occurs
because indices are repeated four times [e.g. $\mu$ in eq.~(\ref{TAdef})].
The explicit form of this diagonal hypercubic rotation is
\begin{equation}
Q(x)\to (\Lambda_{1/2}^{\rm HC} \otimes [\Lambda_{1/2}^{\rm HC}]^*) 
Q([\Lambda^{\rm HC}]^{-1} x) \,,
\end{equation}
with $\Lambda_{1/2}^{HC}$ an element of the spinor representation 
restricted to the hypercubic group, and $\Lambda^{HC}$ the
corresponding member of the vector representation.
Since these transformations involve simultaneous Lorentz and flavor
rotations we refer to the group they form as $SW_{4,\rm diag}$.
As noted above, $S_6^{FF(B)}$ is also invariant under the discrete
flavor group $\Gamma_4$, the generators of which transform as a vector
under $SW_{4,\rm diag}$. Thus the flavor-rotation group of $S_6^{FF(B)}$
is $\Gamma_4\semitimes SW_{4,\rm diag}$.

The symmetry group of $S_6$ is given by the intersection of
the symmetry groups of its components. Combining the above results,
we find that $\Gamma_4\semitimes SW_{4,\rm diag}$ is the symmetry
group in general, and that in the chiral limit we get the additional
$U(1)_A$.
Adding back in fermion number, charge conjugation and parity,
the resulting groups are indeed those of the 
underlying lattice theory~\cite{GS}.
This is most easily seen using the presentation of the lattice group
given in Ref.~\cite{toolkit}: the $\Gamma_4$ is generated
by single-site translations,
while the $SW_{4,diag}$ is  generated by lattice rotations. 

In summary, we have learned that the bulk of flavor symmetry breaking
is caused by the subclass of four-fermion operators in which
flavor and Lorentz indices are correlated. 
We will now show that the contributions of these operators are 
suppressed in low momentum pion interactions.

\section{Effective Chiral Lagrangian}\label{sec:chiral}

In this section we work out the modifications to the chiral Lagrangian
implied by the addition of the action $a^2 S_6$.
We begin by recalling the form of the Lagrangian describing
the long-distance dynamics of the QCD action, $S_4$, close to the chiral limit.
The spontaneous breakdown of $SU(4)_L\times SU(4)_R $ to vector
$ SU(4) $ gives rise to 15 Goldstone modes, described by fields $\phi_i$.
These can be collected into an $SU(4)$ matrix
\begin{equation}
\Sigma(x) = \exp( i \phi / f) \,,\qquad
\phi = \sum_{a=1}^{15} \phi_a T_a \,,
\end{equation}
where $f$ is the pion decay constant (normalized to $f_\pi=132\,$MeV).
We adopt an unconventional normalization for the generators,
\begin{equation}
T_a = \left\{ \xi_\mu, i\xi_{\mu5}, i\xi_{\mu\nu}, \xi_5 \right\} \,.
\end{equation}
Under $SU_L(4) \times SU_R(4)$, $\Sigma$ transforms as 
\begin{equation}
\Sigma(x) \longrightarrow  L \Sigma(x) R^{\dagger}
\end{equation}
where $L(x) \in SU_L(4)$ and $R(x) \in SU_R(4) $.
The chiral Lagrangian, correct up to quadratic order in meson masses
and momenta, is
\begin{eqnarray}
\CL_\chi^4 = \frac{f^2}{8} {\sf Tr} (\partial_\mu \Sigma 
\partial_{\mu} \Sigma^{\dagger} )  - 
\frac{1}{4} \mu\, m \, f^2 {\sf Tr}( \Sigma + \Sigma^{\dagger} ) \,,
\label{effact-1}
\end{eqnarray}
with $\mu$ a constant of $O(\Lambda_{\rm QCD})$.
Expanding this in powers of $\phi$ one finds 15 degenerate pions
with masses given by
\begin{equation}
m_\pi^2 = 2 \mu m \left[1 + O(m/\Lambda_{\rm QCD}) \right] \,.
\label{eq:mpisqchiral}
\end{equation}
The leading order term is the tree-level result,
while the corrections come from loop diagrams
and from higher order terms in the chiral Lagrangian.

The addition of $a^2 S_6$ breaks chiral symmetry and 
lifts the degeneracy of the pions. Generically, eq.~(\ref{eq:mpisqchiral})
becomes 
\begin{equation}
m_\pi^2 = c a^2\Lambda_{\rm QCD}^4 +
2 \mu m \left[1 + O(m/\Lambda_{\rm QCD}) + O(a^2\Lambda_{\rm QCD}^2) \right] 
\end{equation}
where $c$ is a constant of order unity.
Contributions proportional to $a^2$ are due to $S_6$,
and lead to massive pions even in the chiral limit.
The only exception is the pion with flavor $\xi_5$ which remains 
massless because $S_6$ respects the $U(1)_A$ symmetry when $m\to0$.
It is instructive to rewrite the general form as
\begin{equation}
{m_\pi^2 \over \Lambda_{\rm QCD}^2 } =
  a^2\Lambda_{\rm QCD}^2 
+ {m\over\Lambda_{\rm QCD}}
+ a^2\Lambda_{\rm QCD}^2 {m\over\Lambda_{\rm QCD}}
+ {m^2\over\Lambda_{\rm QCD}^2}
+ \dots 
\,,
\label{eq:mpiexp}
\end{equation}
where we have dropped all constants. This shows explicitly that
we are doing a joint expansion in $a^2$ and $m$ about the continuum
and chiral limits. If we treat both parameters as small, then
we can ignore all but the first two terms.
The key observation is then the following.
While the combined effect of all the terms proportional to $a^2$ 
(and any power of $m$) in eq.~(\ref{eq:mpiexp}) is to 
break the chiral and Lorentz symmetries down to the lattice subgroup 
(since $S_6$ itself breaks the symmetry in this way),
it is possible that the $a^2$ term which remains in the chiral
limit has a larger symmetry.
This turns out to be the case, and implies certain exact degeneracies
between pions in different lattice irreps in the chiral limit.

To demonstrate this we need to determine the mapping of the operators in $S_6$
into the chiral Lagrangian.
This is done by matching the transformation
properties of operators under Lorentz and chiral symmetries.
We denote the resulting contributions to the total chiral Lagrangian
$\CL_\chi^6$, i.e.
\begin{equation}
\CL_\chi = \CL_\chi^4 + a^2 \CL_\chi^6 + O(a^4) \,.
\label{effact-2}
\end{equation}
A precise statement of our result is that the part of $\CL_\chi^6$
without derivatives, i.e. the potential $\CV_\chi^6$, is invariant
under flavor $SO(4)$ transformations.\footnote{%
We adopt the sign convention that $\CL_\chi^6=\CV_\chi^6$ plus terms
containing derivatives.}
Terms in $\CL_\chi^6$ involving derivatives do break $SO(4)$ down to 
the lattice symmetry group.
But for very low momentum interactions, such that $p^2 \sim m_\pi^2$,
these terms give contributions suppressed relative to those of 
$\CV_\chi^6$ by $m_\pi^2/\Lambda_{\rm QCD}^2$.
For example in the expression for $m_\pi^2$, eq.~(\ref{eq:mpiexp}),
$\CV_\chi^6$ gives rise to the $a^2$ term on the r.h.s, while the
terms in $\CL_\chi^6$ containing derivatives give rise to the
non-leading term proportional to $a^2 m$ and $a^4$.
Note, however, that for pion interactions with $p^2\sim \Lambda_{\rm QCD}^2$
the $SO(4)$ symmetry will be broken.

In the following we work through the operators in $S_6$ matching them
into $\CL_\chi^6$, with the aim of finding those which contribute to
the potential $\CV_\chi^6$.

\subsection{Matching of operators in $S_6^{\rm glue}$ and $S_6^{\rm bilin}$}

As noted above, most of these operators have the same symmetries as $S_4$,
and in particular do not break the $SU(4)$ flavor symmetry.
They lead to a renormalization of
the terms in the continuum chiral Lagrangian $\CL_\chi^4$
by  a factor of the form $1 + a^2\Lambda_{\rm QCD}^2 + a^2 m^2$.
They do contribute to $\CV_\chi^6$,
but their leading contribution to $m_\pi^2$ is of order $O(a^2 m)$,
and thus of second order in our joint expansion.

There are also terms in $S_6^{\rm glue}$ and $S_6^{\rm bilin}$ 
which are not invariant under rotations.
These map into operators 
which themselves are not rotationally invariant, 
which requires at least four derivatives, e.g.
\begin{equation}
\sum_\mu {\sf Tr} (\partial_\mu^2 \Sigma 
\partial_{\mu}^2 \Sigma^{\dagger} ) \,.
\end{equation}
These lead to rotation non-invariance in the pion propagator,
and thus direction dependence of the extracted pion mass,
but they do not break the flavor symmetry.
Their contributions to $m_\pi^2$ are of size
$a^2 m_\pi^4$ and thus proportional to $a^2 m^2$, $a^4 m$ or
$a^6$, and so are of third order in our joint expansion.

\subsection{Matching of operators in $S_6^{\rm FF(A)}$}

The first flavor-breaking contribution to the potential $\CV_\chi^6$
comes from the operators in $S_6^{\rm FF(A)}$.
These can match onto operators without derivatives because they
are singlets under rotations.
The explicit mapping is worked out in App.~\ref{app:matching}. 
At this stage, the potential retains an $SO(4)$ flavor symmetry,
since this is the symmetry of all the operators in $S_6^{\rm FF(A)}$.

Using the results of App.~\ref{app:matching}, we find that
the contribution to the potential is 
\begin{eqnarray}
- \CV_\chi^6 &=& \hphantom{+\,}
 C_1 \Tr(\xi_5\Sigma \xi_5 \Sigma^\dagger)
\nonumber \\ &&
+\, C_2\, \half 
\left[\Tr(\Sigma^2) -\Tr(\xi_5\Sigma\xi_5\Sigma) + h.c.\right] \,.
\nonumber \\ &&
+\, C_3\, \half \sum_\nu \left[\Tr(\xi_\nu\Sigma\xi_\nu \Sigma) + h.c.\right] 
\nonumber \\ &&
+\, C_4\, \half \sum_\nu 
\left[\Tr(\xi_{\nu5}\Sigma\xi_{5\nu} \Sigma) + h.c.\right] 
\nonumber \\ &&
+\, C_5\, \half \sum_\nu \left[\Tr(\xi_\nu\Sigma \xi_\nu \Sigma^\dagger) 
         - \Tr(\xi_{\nu5}\Sigma \xi_{5\nu} \Sigma^\dagger) \right]
\nonumber \\ &&
+\, C_6 \sum_{\mu<\nu} 
\Tr(\xi_{\mu\nu}\Sigma \xi_{\nu\mu} \Sigma^\dagger) \,.
\label{eq:massterms}
\end{eqnarray}
The six unknown coefficients $C_i$ are all of size $\Lambda_{\rm QCD}^6$.
All terms respect the enlarged flavor symmetry and are Lorentz singlets.
To obtain the result (\ref{eq:massterms}), we have collected
the contributions from the different operators
in eqs.~(\ref{eq:VSmap}-\ref{eq:VTmap}), 
(\ref{eq:SVmap}), (\ref{eq:SAmap}), (\ref{eq:TVmap}), and (\ref{eq:STmap}),
and then used Fierz transformations to convert all operators to a form
containing a single flavor trace.

\subsection{Matching of operators in $S_6^{\rm FF(B)}$}

Our final task is to match the operators in $S_6^{\rm FF(B)}$,
i.e. those in which spin and flavor are correlated.
These are the operators which break flavor down to the lattice subgroup.
The key point in this matching is that all the operators transform  
non-trivially under rotations. 
To represent these operators in the chiral Lagrangian
requires the introduction of derivatives, 
since these are the only objects available to introduce
non-trivial rotation dependence.
This implies that the operators in $S_6^{\rm FF(B)}$ {\em make no
contribution to the potential}.
Thus the result eq.~(\ref{eq:massterms}) is the full result for
$\CV_\chi^6$, and is $SO(4)$ flavor symmetric as claimed.
This is the central result of this paper.

Although complete flavor breaking does not occur at first
order in our joint expansion, it is interesting to see how it
enters at next order. 
An example of the operators that appear when matching
$S_6^{\rm FF(B)}$ is
\begin{equation}
\,[T_\mu\times V_\mu] \longrightarrow
\sum_{\mu\ne\nu\ne\rho\ne\mu} 
\Tr\left((\partial_\mu-\partial_\nu)\Sigma\,\xi_\rho\right)\,
\Tr\left((\partial_\mu-\partial_\nu)\Sigma\,\xi_\rho\right) \,.
\end{equation}
This leads to a direction dependent contribution to 
pion masses that is correlated with their flavors. 
For example, this term renormalizes the
mass-squared of a $\vec p=0$ pion with flavor $\xi_j$ ($j=1-3$) by a factor
of $1+ a^2\Lambda_{\rm QCD}^2$, while it does not effect the mass
of the pion with flavor $\xi_4$.
Since in leading order in our joint expansion, the pion mass-squared is
proportional to $a^2$ and $m$, this example shows that the
second order correction terms proportional to both $a^4$ and $a^2 m$ 
in eq.~(\ref{eq:mpiexp}) break the symmetry down to the lattice subgroup.

\subsection{Contributions of $O(a^4)$}

We close this section with an observation concerning higher order
discretization errors. The numerical results discussed in the
following section indicate that the $SO(4)$ symmetry is 
only weakly broken even at fairly large lattice spacing.
This raises the question of whether higher order discretization errors
break the flavor $SO(4)$. In particular, is the enlarged flavor
symmetry respected by higher order discretization errors in the
chiral limit?

The analysis of the previous subsection has, in fact,
answered this question in the negative.
Terms in the chiral Lagrangian involving derivatives do break $SO(4)$, 
and, through wave-function renormalization,
give flavor-breaking contributions of $O(a^4)$ to $m_\pi^2$.
Nevertheless, it is interesting to know whether the ``direct'' contributions
of $O(a^4)$, i.e. those in the potentials $\CV_\chi^{n\ge8}$, break $SO(4)$.

To study this question,
all we need to know is that,
in the chiral limit, any operator in $\CV_\chi^{n\ge8}$ 
must be symmetric under
\begin{equation}
U(1)_A \times (\Gamma_4 \semitimes SW_{4,\rm diag}) \,,
\label{eq:chirallatsym}
\end{equation}
for this is the symmetry of the lattice theory for $m=0$.
The issue is whether all operators in $\CV_\chi^{n\ge8}$ that respect this
symmetry also respect a larger flavor symmetry 
(as was the case for $\CV_\chi^6$).
That the answer is negative is shown by the example
\begin{equation}
\sum_\nu \Tr\left(
\xi_\nu \Sigma \xi_\nu \Sigma \xi_\nu \Sigma \xi_\nu \Sigma
\right) + h.c.
\label{eq:higherorder}
\end{equation}
This does respect a larger symmetry than (\ref{eq:chirallatsym}),
because (when integrated over all space-time) 
it is invariant under rotations.
But its flavor symmetry is the hypercubic group $SW_4$, and not $SO(4)$.
Thus flavor is completely broken by higher order terms in the
potential. 

Nevertheless, it is amusing to note that the contributions of such
terms to the pion masses do respect $SO(4)$. This is because, when
calculating pion masses, two of the four factors of $\Sigma$ are replaced
by the identity, and the operator ``contracts'' to one which
is $SO(4)$ symmetric.

This analysis does not tell us whether the new term
(\ref{eq:higherorder}) is proportional to $a^4$ or a higher power of $a^2$.
To determine this one must study the quark operators of dimension 8
allowed by the lattice symmetries, a task we have not undertaken.

\section{Results for pion masses}\label{sec:pion}

The partial restoration of continuum symmetries in $\CV_\chi^6$
applies to all the multi-pion interactions encoded in the chiral Lagrangian.
We consider here only the simplest application of this result,
namely to the masses of the pions.

In the continuum, the pions form a 15-plet of flavor $SU(4)$,
and are degenerate. On the lattice, states are classified by the
symmetries of the transfer matrix (a subgroup of the symmetries of
the lattice action). As shown by Golterman \cite{goltmesons}, the pions
fall into 7 irreps of this group: four 3-dimensional reps
with flavors $\xi_i$, $\xi_{i5}$, $\xi_{ij}$ and $\xi_{i4}$,
and three 1-dimensional reps with flavors $\xi_4$, $\xi_{45}$ and $\xi_5$.
Here we have chosen the 4'th direction as Euclidean time.
These results hold irrespective of the quark mass.

Close to both the chiral and continuum limits,
the pion masses are given by
\begin{equation}
M_\pi(T_a)^2 = 2 \mu m + a^2 \Delta(T_a) + O(a^2 m) + O(a^4) \,,
\label{eq:pimassform}
\end{equation}
with $\Delta(T_a) \sim \Lambda_{\rm QCD}^4$ arising from $\CV_\chi^6$.
Since $\CV_\chi^6$ respects flavor $SO(4)$
(as well as a discrete group which plays no r\^ole in relating
pion masses), the pions fall into $SO(4)$ representations:
\begin{itemize}
\item
A one-dimensional irrep with flavor $\xi_5$, which receives no mass
from any of the terms in $\CL_\chi^6$, because of the exact
axial $U(1)$ symmetry: 
\begin{equation}
\Delta(\xi_5) = 0 \,.
\label{eq:delta5}
\end{equation}
\item
Two four-dimensional irreps with flavors $\xi_\mu$ and $\xi_{\mu5}$. These
receive masses
\begin{eqnarray}
\Delta(\xi_\mu) &=& {16 \over f^2} 
(C_1+C_2+C_3+3 C_4+C_5+3 C_6) \,, 
\label{eq:ximu}\\
\Delta(\xi_{\mu5}) &=&{16 \over f^2} 
(C_1+C_2+3 C_3+ C_4-C_5+3 C_6) \,.
\label{eq:ximu5}
\end{eqnarray}
\item
One six-dimensional irrep with flavor $\xi_{\mu\nu}$, 
$\mu \ne \nu$, with mass
\begin{equation}
\Delta(\xi_{\mu\nu}) = {16 \over f^2} (2C_3+2 C_4+4 C_6) 
\label{eq:ximunu}
\end{equation}
\end{itemize}
The mass shifts for these representations are independent, and thus
we can make no predictions for the ordering or splittings.
We also cannot predict the {\em sign} of the shifts, although
numerical data indicates that they are all positive with present
forms of the staggered action.

We can finally see the degeneracies predicted in the chiral limit
at finite lattice spacing. The lattice irreps with flavors
$\xi_i$ and $\xi_4$ join together, as do those with flavors
$\xi_{i5}$ and $\xi_{45}$, and those with flavors
$\xi_{ij}$ and $\xi_{i4}$.

How well do these predictions compare to numerical data?
Two groups have calculated the masses of all seven pion irreps
with errors small enough to allow a test.
Ishizuka {\em et al.} made an extensive study of flavor symmetry
breaking several years ago~\cite{ishizuka}.
They worked in the quenched approximation at $\beta=6.0$,
and used two quark masses $m=0.01$ and $m=0.02$.
These masses corresponding roughly to $m_s/3$ and $2 m_s/3$, 
with $m_s$ the physical strange quark mass, and thus are moderately small.
The chiral expansion parameter $(m_\pi/4\pi f_\pi)^2$ 
is about $25\%$ for such masses.
The second calculation has been done recently
by Orginos and Toussaint~\cite{orginos}.
They have done extensive calculations using
a variety of perturbatively improved staggered fermion actions
using moderately light dynamical quarks. 

\begin{figure}[tb]
\vspace{-0.1truein}
  \centerline{\psfig{file=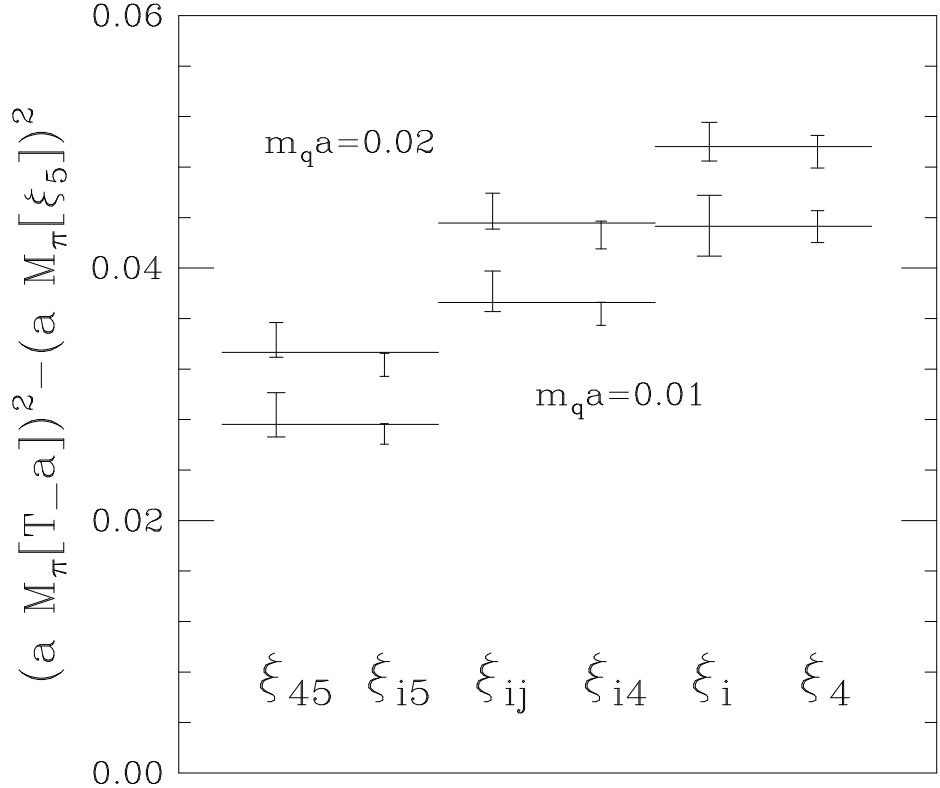,height=5truein}}
\vspace{-0.5truein}
\caption{Splittings between non-Goldstone and Goldstone pions
at $\beta=6$ in the quenched approximation~\protect\cite{ishizuka}.
Results are in lattice units. The solid horizontal lines show
the average value for the three pairs which are predicted to
become degenerate in the chiral limit. Errors are approximate,
as they have been obtained ignoring the error in the Goldstone pion mass,
and the correlation between the masses of pions in different representations.}
\label{fig:ishizuka}
\end{figure}

\begin{figure}[tb]
\vspace{-0.1truein}
  \centerline{\psfig{file=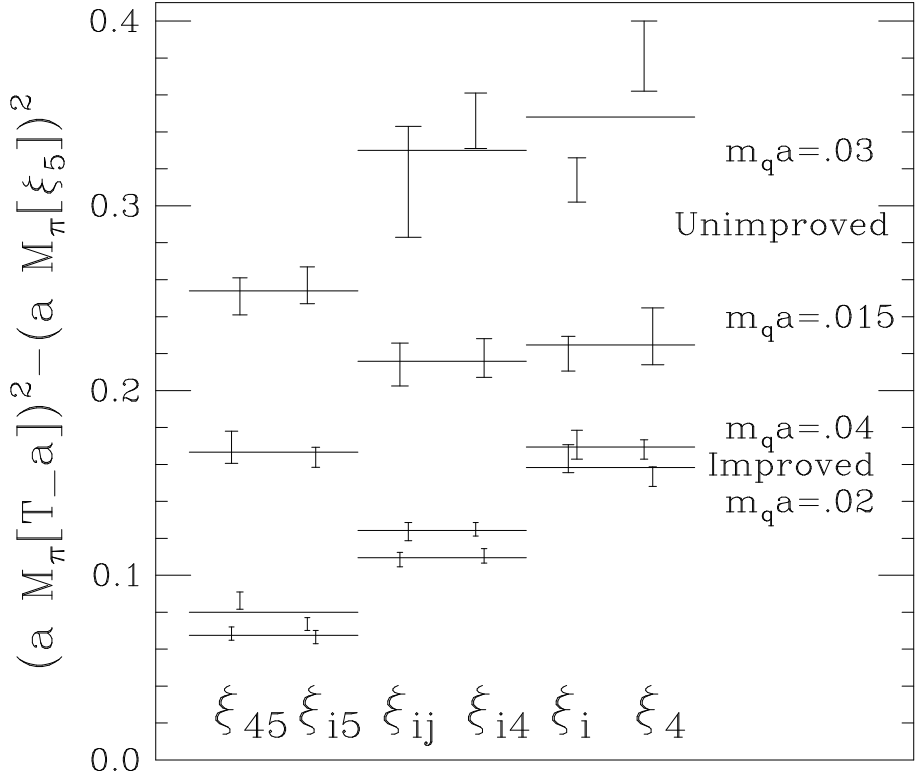,height=5truein}}
\vspace{-0.5truein}
\caption{Splittings between non-Goldstone and Goldstone pions
in dynamical simulations~\protect\cite{orginos}. 
Notation is as in Fig.~\ref{fig:ishizuka}.
The upper two sets of points come from simulations
with unimproved gauge and fermion actions, while the lower two
use improved gauge and fermion actions (the latter being the ``fat+Naik''
action). For clarity, some points have been offset horizontally.} 
\label{fig:orginos}
\end{figure}

It turns out that the predicted degeneracies are observed in almost
all the data sets, even at the largest quark masses.
This is illustrated in Figs.~\ref{fig:ishizuka} and
\ref{fig:orginos}, which show, respectively, 
the results of Ref.~\cite{ishizuka} and a subset of the results
from Ref.~\cite{orginos}. For each of the six non-Goldstone pion irreps,
we plot the difference
\begin{equation}
a^2 \left[M_\pi(T_a)^2 - M_\pi(\xi_5)^2\right]
= a^4 \Delta(T_a)\,\left[1 + O(m/ \Lambda_{\rm QCD}) + O(a^2 \Lambda_{\rm
QCD}^2) \right] \,,
\label{eq:massdiffdef}
\end{equation}
which removes the common contribution proportional to the quark mass.
We order the flavors so that pairs which are predicted to
become degenerate in the chiral limit are adjacent.
For all pairs except one ($\xi_i$ and $\xi_4$ at the heaviest
quark mass with unimproved fermions in Fig.~\ref{fig:orginos}),
degeneracy is observed within errors.\footnote{%
In simulations, the masses of the various pions are correlated,
and thus the differences between them are more significant than
the error bars suggest. It would be interesting to directly
calculate the errors on these differences.}
Furthermore, there is, in most cases, a statistically
significant difference between the masses of the pions in
different $SO(4)$ representations.

The numerical data are thus consistent with a flavor $SO(4)$ symmetry,
not only in the chiral limit but also for non-zero quark masses.
As noted in the previous section, this symmetry is broken
by both the higher order terms in eq.~(\ref{eq:massdiffdef}).
In particular, $SO(4)$ breaking by the correction proportional to 
$m$ would be twice as large for the heavier masses in each of the data sets.
It is apparent from the figures that such $SO(4)$ breaking is small,
and thus that the bulk of this correction is $SO(4)$ symmetric.
This result is not explained by the chiral Lagrangian analysis.

It is interesting to compare the two data sets in Fig.~\ref{fig:orginos}.
The two sets are approximately matched in the following sense:
for both the smaller and the larger quark masses,
the values of $a M_\rho$ and $a M_\pi(\xi_5)$ from the
unimproved and improved simulations agree closely.
This means that they have roughly the same lattice spacing and quark mass,
and can thus be used to observe the effect of improvement 
on flavor symmetry breaking.
At non-zero quark mass the flavor breaking is substantially reduced,
particularly for the pions with flavor $\xi_{\mu5}$.
On the other hand, if one extrapolates linearly to the chiral limit,
the effect of improvement is much smaller.
It is also noteworthy that, for improved fermions, for which one
can do the chiral extrapolation with little uncertainty,
the pattern of flavor breaking is consistent with only the
$C_4$ term in eqs.~(\ref{eq:ximu}-\ref{eq:ximunu}) being substantial.
We have no explanation for this observation.

Finally, we can use the data to look at the absolute size of the
discretization errors in the chiral limit.
The quantity shown in the Figures should be 
of size $(a\Lambda)^4$, with $\Lambda\sim \Lambda_{\rm QCD}$.
Performing a linear extrapolation to $m=0$, and taking the
result for the flavors $\xi_{\mu\nu}$ as representative,
we find the correction to be $0.031$ for the quenched data,
and $0.096$ (resp. $0.11$) for the improved (resp. unimproved) 
dynamical simulations.
The lattice spacing for these simulations (obtained from
extrapolating the rho mass to the chiral limit) is roughly
$1/a=2\,$GeV for the quenched data, and $1.3\,$GeV for both
dynamical simulations. Thus one finds that the scale describing
the discretization errors is $\Lambda\approx 0.8$, $0.7$ and
$0.7\,$GeV for the three cases. These rather large values 
illustrate the need for further improvement of staggered fermions.

\section{Non-Goldstone particles}\label{sec:nonGoldstone}

An interesting question is whether the partial symmetry 
restoration found in the pion sector extends to other
particles such as the vector mesons and light baryons. 
A naive argument against such an extension goes as follows.
Even for pions, the enlarged symmetry is not respected by
terms proportional to $a^2 p^2$. For other hadrons, e.g. the
$\rho$ meson, one has $|p^2| = m_\rho^2 \gg m_\pi^2$, and so
the flavor-breaking terms are not suppressed compared to other
discretization errors.
We have studied this question in detail for vector mesons,
and find, as explained in this section,
that indeed symmetry restoration does not occur.

The essential difference from pions is that the vectors remain
massive in the combined chiral and continuum limits.
Indeed, the simplest way to include them in the chiral Lagrangian
is by expanding in inverse powers of their masses~\cite{JMW}.
At leading order they are sources having fixed velocity.
Higher order terms fit naturally into a chiral expansion because
the expansion parameters match: $1/m_\rho\approx 1/(4 \pi f_\pi)$.

The key point is that there is no obstacle to constructing terms,
 at leading order in the chiral expansion, with the same transformation 
properties as those contained in $\CL_6^{FF(B)}$. 
In particular, unlike for pions, one does not have to
build Lorentz non-singlets using derivatives. Instead, one
has two new Lorentz vectors, 
namely the rho field itself $\rho_\mu$ and
the four-velocity of the heavy source, $v_\mu$.
What is particularly important is that additional factors of $v_\mu$ 
can be added at no cost in the chiral expansion. 
To see how this works in detail is straightforward but tedious.
We give only an example. The result of matching the operators
$([V\times V]-[A\times A])$ and
$([V_\mu\times V_\mu]-[A_\mu\times A_\mu])$ 
onto operators in the chiral Lagrangian includes the following mass terms,
\begin{equation}
\sum_{\mu,\nu}\Tr[\rho_\mu^\dagger \xi_\nu \rho_\mu \xi_\nu] +
\sum_\nu \Tr[\rho_\nu^\dagger \xi_\nu \rho_\nu \xi_\nu] +
\sum_{\mu,\nu}  
\Tr[\rho_\mu^\dagger \xi_\nu \rho_\mu \xi_\nu] v_\nu v_\nu \,,
\end{equation}
with unknown coefficients. Here $\rho_\mu$ is a flavor matrix containing
the $SU(4)$ 15-plet of vector mesons.
There are also similar terms with ``axial'' flavor.
Note that both the second and third terms break Lorentz invariance
and flavor symmetry.

Performing a similar analysis with all the operators in $\CL_6$ we
find the following form for the $a^2$ contribution to the
vector meson mass matrix in the chiral limit:
	\begin{eqnarray}
	{\cal L}^6_\rho &=& R_1 \sum_{\mu} \Tr [ \rho_\mu^\dagger \rho_\mu ] 
		+ R_2 \sum_{\mu} 
		\Tr [ \rho_\mu^\dagger \xi_5 \rho_\mu \xi_5 ]		
	\nonumber \\ & &
		+ R_3 \sum_{\mu,\nu}
		\Tr [ \rho_\mu^\dagger \xi_{\nu} \rho_\mu \xi_{\nu} ]
		+ R_4 \sum_{\mu,\nu}
		\Tr [ \rho_\mu^\dagger \xi_{\nu5} \rho_\mu \xi_{\nu5} ]
		+ R_5 \sum_{\mu,\nu < \lambda}
		\Tr [ \rho_\mu^\dagger \xi_{\nu\lambda} 
			\rho_\mu \xi_{\nu\lambda} ]		
	\nonumber \\ & &
		+ R_6 \sum_{\mu}
		\Tr [ \rho_\mu^\dagger \xi_{\mu} \rho_\mu \xi_{\mu} ]
		+ R_7 \sum_{\mu}
		\Tr [ \rho_\mu^\dagger \xi_{\mu5} \rho_\mu \xi_{\mu5} ]	
		+ R_8 \sum_{\mu \ne \nu}
		\Tr [ \rho_\mu^\dagger \xi_{\mu\nu} 
			\rho_\mu \xi_{\mu\nu} ]		
	\nonumber \\ & &
                + R_9 \sum_{\mu}
                \Tr [ \rho_\mu^\dagger \xi_4 \rho_\mu \xi_{4} ]
                + R_{10} \sum_{\mu}
                \Tr [ \rho_\mu^\dagger \xi_{45} \rho_\mu \xi_{45} ]
                + R_{11} \sum_{\mu,\nu}
                \Tr [ \rho_\mu^\dagger \xi_{4\nu}
                        \rho_\mu \xi_{4\nu} ] \,.
	\label{eq:L6rho}
	\end{eqnarray}
Here we have specialized to the rest frame, in which $\vec v=0$
and $\rho_4=0$.
The unknown coefficients $R_i$ are of size $a^2\Lambda_{\rm QCD}^4$.
These eleven coefficients break the vector meson octet down into
the eleven lattice irreps found by Golterman~\cite{goltmesons}. 
These are the states created by the bilinears 
	\begin{eqnarray}
	\bar Q (\gamma_k \otimes \xi_5) Q, &&
	\bar Q (\gamma_k \otimes \xi_m) Q,\   
	\bar Q (\gamma_k \otimes \xi_4) Q,
 	\nonumber \\
	\bar Q (\gamma_k \otimes \xi_k) Q, &&
	\bar Q (\gamma_k \otimes \xi_{m5}) Q, \
	 \bar Q (\gamma_k \otimes \xi_{45}) Q, 
	\nonumber \\
	\bar Q (\gamma_k \otimes \xi_{k5}) Q,  &&
	\bar Q (\gamma_k \otimes \xi_{m4}) Q,\
        \bar Q (\gamma_k \otimes \xi_{lm} ) Q,
	\nonumber \\
	\bar Q (\gamma_k \otimes \xi_{k4} ) Q, &&
	\bar Q (\gamma_k \otimes \xi_{mk}) Q,
	\end{eqnarray}
where $ k \ne l \ne m $.
Thus there are no predicted degeneracies. Note that the
terms in the last line of eq.~(\ref{eq:L6rho}), which are present because
of the vector $v_\mu$, are necessary to lift all degeneracies.

The conclusion we draw from this example is that the pions are
a special case, and that for all other particles
there is no reason to expect a partial restoration of the lattice symmetry.
At the present time there are no numerical results accurate or
extensive enough to test this prediction.

\section{An ``Aoki-phase'' for staggered fermions?}\label{sec:phase}

As the quark mass is lowered, there comes a point at which the
effects of the quark mass term and the leading discretization errors
are comparable. This occurs for physical masses $m \sim a^2$, 
implying lattice masses $a m \sim a^3$.
At this point the competition between the two contributions can lead to
unexpected patterns of symmetry breaking.
For Wilson fermions one has the possibility of
spontaneous breakdown of flavor and parity symmetries,
i.e. an ``Aoki-phase'' containing exactly massless Goldstone 
bosons~\cite{aoki}.
What we wish to point out here is that
an analogous phenomenon can occur with staggered fermions---the
$SO(4)$ flavor symmetry of the $a^2$ terms can be broken spontaneously.
The method we use is a direct generalization of the work of
Ref.~\cite{ss} on the Aoki-phase for Wilson fermions.

That symmetry breaking can occur is clear from the form for
pion masses, eq.~(\ref{eq:pimassform}).
If the corrections of $O(a^2)$ (called $\Delta(T_a)$ in the equation)
are negative for some pion flavors, then $M_\pi(T_a)^2$ will 
become negative for small enough quark mass. This signals vacuum
instability and the possibility of spontaneous symmetry breaking.
As noted in Sec.~\ref{sec:pion}, present versions of the staggered
action appear to have all $\Delta(T_a)$ positive, in which case 
symmetry breaking does not occur.\footnote{%
When $m$ becomes negative,
the vacuum expectation value $\langle \Sigma\rangle$ flips from
$+1$ to $-1$, and the result (\ref{eq:pimassform}) remains valid
with $m\to |m|$.
}
Nevertheless, it is possible that alternate improved versions will lead
to symmetry breaking, and we discuss this possibility briefly.

Before doing so, however, we stress an important 
difference between the Aoki-phase with Wilson fermions 
and the present analysis.
With Wilson fermions, the flavor symmetry in question is an exact 
symmetry of the lattice theory, and its breaking leads to massless 
Goldstone bosons at non-zero lattice spacing.
In the present case, the flavor symmetry is only approximate---it
is violated by terms of size $a^4$.
These terms lead only to small corrections to the masses of most 
of the pions in the region of interest ($M_\pi^2\sim a^2\Lambda^4$).
The only exceptions are the Goldstone pions ($\phi_G$ below), which
become pseudo-Goldstone bosons with $M_G^2\sim a^4\Lambda^6$.

To simplify the discussion we assume that only the $C_4$ term
in ${\cal L}_\chi^{6}$ is present [see eq.~(\ref{eq:massterms})],
and assume that $C_4$ itself is negative. 
A straightforward calculation finds the following features as the
magnitude of the quark mass is reduced.
For $|m|$ greater than the critical value
$m_c = 24 a^2 |C_4| / (\mu f^2) \sim a^2\Lambda^3$,
the condensate is given by $\langle \Sigma\rangle= {\rm sign}(m) 1 $,
and the mass formulae of Sec.~\ref{sec:pion} apply with $m\to |m|$.
When $|m|=m_c$ the pions with flavor $\xi_\mu$ become massless
[see eqs~.(\ref{eq:delta5})-(\ref{eq:ximunu})],
and there is a second order phase transition.
For $|m|\le m_c$ it is useful to introduce the angle 
$\theta$ defined by $\cos\theta=m/m_c$, with $0\le \theta \le \pi$.
Minimizing the potential, one finds that
the condensate swings from $+1$ to $-1$ according to
\begin{equation}
\langle \Sigma \rangle = \exp(i \theta [\xi_\mu \cdot n_\mu]) 
= \cos\theta + i \sin\theta [\xi_\mu \cdot n_\mu]\,,
\end{equation}
where $n_\mu$ is an arbitrary vector of unit length.
For $0<\theta < \pi$, 
flavor $SO(4)$ is broken to an $SO(3)$ subgroup,
and there are three Goldstone pions.
As noted above,
these are not exactly massless because terms in ${\cal L}_\chi$
of $O(a^2 m)$ and $O(a^4)$ break the $SO(4)$ explicitly.

It is instructive to work out the full spectrum of pions in
the broken phase. We display these for $n_\mu = \delta_{\mu 4}$,
i.e. for the case in which the condensate is
$\langle \Sigma \rangle = \exp(i \theta \xi_4)$.
The pions are labeled by their flavor, defined by the
direction of excitation of the condensate,
$\Sigma = \langle \Sigma \rangle \exp(i \phi_a T_a)$.
The results are given in units of $16 a^2 |C_4|/f^2$.
\begin{itemize}
\item
Flavor $T_a=\xi_j,i\xi_{4j}$, with $j=1,2,3$. For each $j$, these two
flavors mix, and there are two mass eigenstates:
\begin{eqnarray}
\phi_G = \cos\theta \phi(\xi_j) + \sin\theta \phi(\xi_{4j}) \,,&&
\qquad M_G^2 = 0 \,;\\
\phi_{NG1} = \cos\theta \phi(\xi_{4j}) - \sin\theta\phi(\xi_{j}) \,,&&
\qquad M_{NG1}^2 = 1 \,.
\end{eqnarray}
At the end-points, the Goldstone modes are pure flavor $\xi_j$,
but at the mid-point they become pure flavor $\xi_{4j}$.
\item
Flavor $T_a=\xi_4$. This single state has
\begin{equation}
M^2= 3 \sin^2\theta\,.
\end{equation}
This vanishes at the end-points, due to the restoration of $SO(4)$ symmetry,
which makes the masses of all the flavor $\xi_\mu$ states equal.
\item
Flavor $T_a=i\xi_{jk}$. Three states with 
\begin{equation}
M^2= \cos^2\theta\,.
\end{equation}
These become degenerate with the other tensors at the end-points,
due to $SO(4)$ restoration.
This degeneracy recurs at the mid-point, where all six tensors are
massless. This is not, however, caused by $SO(4)$ restoration: 
the $\xi_{4j}$ states are massless due to the 
spontaneous breaking of flavor $SO(4)$,
while the $\xi_{jk}$ states are massless because of an accidental
{\em axial} $SO(3)$ symmetry of the potential about its classical minimum
at $m=0$.
The latter states will presumably become massive when quantum corrections
are included.
\item
Flavors $T_a=i\xi_{j5}$. Three states which are always massive:
\begin{equation}
M^2= 1 + \cos^2\theta\,.
\end{equation}
These become degenerate with the $\xi_j$ at the mid-point, a result
that can be understood as due to the restoration of the 
$U(1)_A$ symmetry.
\item
Flavors $T_a=\xi_5,i\xi_{45}$. These mix according to
\begin{eqnarray}
\phi_{NG2} = \cos\theta \phi(\xi_{45}) + \sin\theta \phi(\xi_{5}) \,,&&
\qquad M_{NG2}^2 = 2 \,;\\
\phi_{NG3} = \cos\theta \phi(\xi_5) - \sin\theta\phi(\xi_{45}) \,,&&
\qquad M_{NG3}^2 = 3 \,.
\end{eqnarray}
The degeneracy between flavors $\xi_{45}$ and $\xi_4$ at the mid-point
is again due to $U(1)_A$ restoration.
\end{itemize}

\section{Conclusions}\label{sec:conc}

We have studied discretization errors for staggered fermions,
by constructing a sequence of effective Lagrangians.
Our central result is that, close to the continuum limit,
the potential term in the chiral Lagrangian describing pion interactions
respects a chiral symmetry group, $U(1)_A\times [\Gamma_4\semitimes SO(4)]$,
which is much larger than that of the underlying lattice theory.
Full $SO(4)$ rotation invariance is also maintained.
The enlarged symmetry predicts degeneracies in the pion masses which
are observed in numerical data with surprisingly small deviations.

While this result is pleasing, it has limited applicability.
The enlarged symmetry is broken in pion interactions by
terms of $O(a^4)$, $O(a^2 m)$ and $O(a^2 p^2)$,
and also does not hold for hadrons other than pions.
We do not know whether it holds for other pion properties,
e.g. the decay constants.

An interesting question is whether our result can be used to simplify
the improvement of the staggered fermion action.
Complete non-perturbative improvement at $O(a^2)$ is prohibitively difficult
because of the number of additional operators that are needed.
If flavor symmetry were partially restored, 
if only approximately as at finite quark mass, 
then one should be able to get away with fewer operators.
An extreme example is the flavor breaking in pion masses alone.
After symmetry restoration there are only three independent differences, 
so one could imagine a non-perturbative tuning of the coefficients of
just three operators so as to eliminate these differences.
This approach was suggested by Lepage~\cite{lepageimp}, 
who pointed out that tree level improvement of the staggered
action contains only three-flavor breaking terms
(which can be written either as four-fermion operators~\cite{luo2}
or as bilinears with fat-links~\cite{lepageimp}).
One can show that these three operators
do give independent contributions to the pion splittings,
so that such tuning is feasible.
On the other hand, the flavor-breaking part of the
potential $\CV_\chi^6$ contains six independent parameters
($C_1-C_6$ in our notation), and so to remove it non-perturbatively
would require introducing six improvement operators.
Furthermore, there is no simplification if one wishes to completely 
improve the full spectrum, for which there is no symmetry restoration.
Thus it is unclear to what extent the limited flavor restoration will help.
Numerical tests are needed to investigate this issue,
and exploratory work has been done
in Refs.~\cite{toussaint,lagae,orginos}.

\section*{Acknowledgments}
We thank Maarten Golterman, Peter Lepage and Doug Toussaint
for helpful discussions and comments.
This work was supported in part by DOE contracts DE-FG03-96ER40956 
and DOE-W7405-ENG-86. 
SS is very grateful to the Center for Computational Physics at the University
of Tsukuba for the hospitality received there while part of this
work was done.

\appendix
\section{Enumerating operators of dimension six}\label{app:ops}

In this appendix we explain how to determine all operators of dimension six
which are singlets under the lattice symmetry group. 
These are the operators which must be added to remove
$O(a^2)$ errors from the staggered action (modulo possible redundancies).
Conversely, when converted to continuum form,
they are the operators which appear in the continuum
effective Lagrangian at $O(a^2)$.
Luo has cataloged these operators previously~\cite{luo2}, 
but did not find all of them.

We use the notation of Refs.~\cite{DS,PS,SP}, 
the relevant parts of which we briefly review here.
The continuum theory describes four fermion flavors, 
which are collected into a field $Q_{\alpha,a}$, with $\alpha$ and
$a$ respectively spinor and flavor indices.
Bilinears are labeled by a spin and a flavor matrix
\begin{equation}
\bar Q_{\alpha,a} (\gamma_S)_{\alpha \beta} (\xi_F)_{a b} Q_{\beta,b} 
\equiv
\bar Q_{\alpha,a} (\gamma_S\otimes\xi_F)_{\alpha a,\beta b} Q_{\beta,b} 
\equiv
\bar Q (\gamma_S\otimes\xi_F) Q \,.
\label{eq:contbilin}
\end{equation}
The spin matrices are labeled by a hypercube vector $S_\mu$, with components
$0$ or $1$,
\begin{equation}
\gamma_S = \gamma_1^{S_1} \gamma_2^{S_2} \gamma_3^{S_3} \gamma_4^{S_4} \,.
\end{equation}
The flavor matrices are labeled similarly by the hypercube
vector $F_\mu$, except that they are built out of the complex conjugate
matrices, $\xi_\mu = \gamma_\mu^*$. This is simply a convention, since
these two bases are unitarily equivalent.
Finally, we use abbreviations such as 
\begin{equation}
\gamma_{\mu\nu}\equiv\gamma_\mu\gamma_\nu \,,\qquad
\gamma_{\mu5}\equiv\gamma_\mu\gamma_5 \,.
\end{equation}

The lattice bilinears which correspond in the continuum limit to
(\ref{eq:contbilin}) are
\begin{equation}
\sum_{CD} \bar\chi(y)_C \sfno{S}{F}_{CD} \chi(y)_D \,,
\end{equation}
where $y$ labels the $2^4$ hypercubes, $C$ and $D$ are
hypercube vectors, and the hypercube field is defined in terms of
the underlying staggered fermion by
\begin{equation}
\chi(y)_C = \frac14 \chi(y+C) \,.
\end{equation}
The matrices $\sfno{S}{F}$ are unitarily equivalent to 
$(\gamma_S\otimes\xi_F)$.

It will be useful in the following to define hypercube fields at
zero physical momentum, which are obtained by averaging over all 
$N_y$ hypercubes:
\begin{equation}
\chi_C = \frac1{N_y} \sum_y \chi(y)_C \,.
\end{equation}
The 16 $\chi_C$ transform
in the defining representation of the lattice symmetry group.
We always use the argument $y$ when referring to the fields which live
on individual hypercubes, and drop this argument for the zero-momentum
fields.

We begin with a comment on fermion bilinears of dimension six.
Luo lists seven such operators. We note however that one linear
combination, which in Luo's notation is
\begin{equation}
\bar\chi(y) (\CD^2 \slash{\CD} - \slash{\CD}\CD^2) \chi(y)
\end{equation}
has negative parity under lattice charge conjugation and
is thus forbidden. 
Furthermore, there are two additional operators, 
not included explicitly by Luo,
\begin{equation}
m^2 \bar\chi(y) \slash\CD \chi(y) \quad {\rm and} \quad
m^3 \bar\chi(y)\chi(y)\,.
\end{equation}
These are, however, redundant, since they can be absorbed 
by changing the normalization of the fields and quark mass.

We now turn to four-fermion operators. We want to construct all such
lattice operators which are singlets under the lattice symmetry group
and which correspond to operators of dimension six in the continuum.
The latter requirement excludes operators containing 
derivatives or factors of the quark mass.
Since the quark mass does not appear, the construction is effectively
in the chiral limit, and thus the lattice operators must also
be singlets under $U(1)_A$.

We construct operators by the following steps. These make use of the
fact that, to generate the lattice symmetry group,
one can replace rotations about a point on the lattice with
those about the center of a hypercube, and similarly for reflections.
These two choices differ only by translations.
\begin{enumerate}
\item
Multiply two bilinears residing on the same hypercube, and sum over
hypercubes.
\begin{equation}
\CO(S,F,S',F') = \sum_y \bar\chi(y) \sfno{S}{F} \chi(y)\,
\bar\chi(y) \sfno{S'}{F'}\chi(y)\,.
\end{equation}
This operator is manifestly invariant under translations by two units.
Gauge invariance is maintained (in a way which maintains rotation and
reflection properties) by including the usual average of products
of gauge links along the shortest paths between the $\bar\chi$ and $\chi$.
This can be done in two ways: joining the $\bar\chi$ and $\chi$
within the bilinears, or between the bilinears.
If we ignore the gauge links, these two choices are related by a Fierz
transformation. This remains true in the presence of the links, 
up to additional higher dimensional operators involving gauge fields.
Thus we need consider only one linear combination of color structures.
We do not need to specify our choice for the following analysis.
Luo uses the linear combination which arises in perturbation theory
\begin{equation}
\half \hbox{\rm (contract within bilinears)} 
- \sixth \hbox{\rm (contract between bilinears)} \,.
\end{equation}
When we compare our operators to his, we are implicitly using this choice.
\item
Form linear combinations of the $\CO(S,F,S',F')$ which are singlets
under the symmetry group of the hypercube---rotations
and reflections---and also charge conjugation. 
We label this group $W_4^{C}$.
The transformation properties of bilinears under 
$W_4^C$ have been worked out by 
Verstegen~\cite{verstegen} and are listed in his Tables 3 and 4.
The bilinears fall into 56 irreps.
To form a four-fermion operator which is a singlet one must combine
two bilinears which live in the same irrep.
Since most $W_4^C$ irreps appear multiple times, there are a large number
of ways of doing this.
\item
At this stage the operators are not invariant under single site translations.
This can be accomplished by applying the projector
\begin{equation}
\prod_\mu \half(1 + \CS_\mu)\,,
\label{eq:transproj}
\end{equation}
with ${\CS}_\mu$ the translation operator in the $\mu$'th direction.
Note that the ${\CS}_\mu$ commute when acting on operators 
with even fermion number, such as the four-fermion operators under
consideration. Thus the order of the factors in the product
in (\ref{eq:transproj})
is unimportant.
\item
The resulting operators are singlets under all lattice symmetries
except $U(1)_A$. We now apply the constraint that they do not contain
derivatives. This can be done by taking the tree level matrix elements
with all external quark states having zero {\em physical} momentum,
i.e. ${\rm mod}(p_\mu a,\pi) = 0$.
This matrix element will vanish if the lattice operator corresponds 
to a continuum operator containing one or more derivatives.
But in evaluating this matrix element each of the hypercube fields can
be replaced by its zero-momentum counterparts: $\chi(y)_A\to\chi_A$.
Thus we are led to consider linear combinations 
of operators of the form
\begin{equation}
\CO'(S,S',F,F') =
\prod_\mu \half(1 + \CS_\mu) \bar\chi \sfno{S}{F} \chi 
\, \bar\chi \sfno{S'}{F'} \chi
\end{equation}
We can simplify this using the known translation properties of
the zero-momentum hypercube field
\begin{equation}
\CS_\mu \chi = \ixi\mu \chi \quad \Rightarrow \quad
\CS_\mu \bar\chi \sfno{S}{F} \chi 
= (-1)^{\tilde F_\mu}\bar\chi \sfno{S}{F} \chi \,,
\label{eq:lattrans}
\end{equation}
where
\begin{equation}
\tilde F_\mu = {\rm mod}(\sum_{\nu\ne\mu} F_\nu,2) \,.
\end{equation}
We thus find that
\begin{equation}
\CO'(S,S',F,F') =
\prod_\mu \half(1 + (-)^{(\tilde F-\tilde{F'})_\mu})
\bar\chi \sfno{S}{F} \chi \,\bar\chi \sfno{S'}{F'} \chi\,.
\end{equation}
This operator vanishes unless $\tilde F=\tilde{F'}$,
which implies $F=F'$. In other words, if the two bilinears have
different flavor, then the projection onto a translation singlet
produces a continuum operator containing derivatives.
Thus we conclude that, in step 2, we must keep only those bilinears
in which $F=F'$. 
\item
Finally, we select those from the resulting list which are singlets
under $U(1)_A$. 
\end{enumerate}
Our claim is that this procedure produces all singlet dimension-6 operators.
\medskip

Thus we must determine which of the lattice four-fermion operators
satisfying $F=F'$ (``diagonal in flavor'') are singlets under hypercube
rotations and reflections. This is a straightforward, though tedious,
group theoretical exercise which can be done using the tools presented
in \cite{verstegen}. First, one finds that to make a singlet it is
necessary that $S=S'$, so that the operator is diagonal in both
spin and flavor.
Then the problem reduces to picking out the $W_4$ singlets in the
product of vectors in which each index appears twice,
e.g. $\gamma_\mu\otimes\gamma_\mu\otimes \xi_\nu\otimes\xi_\nu$. Note
that the ``square'' of a flavor matrix transforms in the same way as
the ``square'' of a spin matrix.
The result is that there are 35 singlets \cite{SP}:
\begin{enumerate}
\item[A)]
There are 25 operators in which the spin and flavor indices are
separately contracted. For the spin matrices we use the notation
\begin{eqnarray}
&&S = 1\otimes 1\,, \quad
P = \gamma_5\otimes \gamma_5\,,\quad
V = \sum_\mu \gamma_\mu \otimes \gamma_\mu \,,\nonumber\\
&&A = \sum_\mu \gamma_{\mu5} \otimes \gamma_{5\mu}\,,\quad
T = \sum_{\mu<\nu} \gamma_{\mu\nu} \otimes  \gamma_{\nu\mu} \,,
\end{eqnarray}
while for flavor matrices we use the same notation with 
$\gamma_\mu \to \xi_\mu$.
The 25 operators are simply the products of the 5 possible spin structures
with the 5 flavor structures.
The notation we use is exemplified by
\begin{equation}
[A\times T] \equiv
\sum_{\mu}\sum_{\mu<\rho}
\bar\chi(y) \sfno{\mu5}{\nu\rho} \chi(y)
\,\bar\chi(y) \sfno{5\mu}{\rho\nu} \chi(y)
\,.
\end{equation}
\item[B)]
The remaining 10 operators have the spin and flavor matrices coupled.
They are
\begin{equation}
[V_\mu\times V_\mu] \equiv \sum_\mu
\bar\chi(y) \overline{(\gamma_\mu\otimes\xi_\mu)} \chi(y)
\,\bar\chi(y) \overline{(\gamma_\mu\otimes\xi_\mu)} \chi(y)
\,,
\label{eq:VVdef}
\end{equation}
along with $[V_\mu\times A_\mu]$, $[A_\mu\times V_\mu]$ and 
$[A_\mu\times A_\mu]$ defined analogously;
\begin{eqnarray}
[V_\mu\times T_\mu] &\equiv& \sum_{\mu<\nu}
\bar\chi(y) \sfno{\mu}{\mu\nu} \chi(y)
\,\bar\chi(y) \sfno{\mu}{\nu\mu} \chi(y) \nonumber\\
&&\mbox{}-
\bar\chi(y) \sfno{\mu}{\mu\nu5} \chi(y)
\,\bar\chi(y) \sfno{\mu}{5\nu\mu} \chi(y)
\,,
\label{eq:VTdef}
\end{eqnarray}
with $[T_\mu\times V_\mu]$, $[A_\mu\times T_\mu]$ and $[T_\mu\times A_\mu]$
defined analogously; and finally $[T_+\times T_+]$ and $[T_-\times T_-]$,
the definitions of which we do not reproduce since
they are $U(1)_A$ non-singlets.
\end{enumerate}

Next we select from these operators those linear combinations
that are $U(1)_A$ singlets. This is automatically true for those
operators consisting of odd bilinears.\footnote{An even/odd bilinear
is one in which the $\bar\chi$ and $\chi$ 
are separated by an even/odd number of links.}
There are 16 of these, all of which were found by Luo.
The relation to Luo's basis is:
\begin{eqnarray}
\,[S\times A] &=& \CF_{9}  \,,\quad [S\times V] = \CF_{13} \,,\\ 
\,[A\times S] &=& -\CF_{3}  \,,\quad [V\times S] = \CF_{11} \,,\\
\,[P\times V] &=& \CF_{10} \,,\quad [P\times A] = -\CF_{17} \,,\\
\,[V\times P] &=& \CF_{4}  \,,\quad [A\times P] = -\CF_{15} \,,\\
\,[T\times V] &=& -(\half\CF_{5}+\CF_{16}) \,,\quad 
\,[T\times A] = (\CF_6 +\half \CF_{12}) \,,\\
\,[V\times T] &=& -(\half\CF_{7}+\CF_{18}) \,,\quad 
\,[A\times T] = (\CF_8+\half\CF_{14}) \,,\\
\,[T_\mu\times V_\mu] &=& (-\half\CF_{5}+\CF_{16}) \,,\quad 
\,[T_\mu\times A_\mu] = (-\CF_6 +\half \CF_{12}) \,,\\
\,[V_\mu\times T_\mu] &=& (-\half\CF_{7}+\CF_{18}) \,,\quad 
\,[A_\mu\times T_\mu] = (-\CF_8+\half\CF_{14}) \,.
\end{eqnarray}
To determine which of the operators composed of even bilinears
are $U(1)_A$ invariant,
we note that any such operator must, 
under flavor-spin Fierz transformation, transform into an operator
composed of two odd bilinears. Conversely, if we Fierz transform each of the
above 16 operators, we will find all even-even diagonal operators.
Note that some of the odd-odd operators Fierz transform back into
odd-odd operators, so that we can end up with less than 16 even-even
operators. Using the Fierz tables collected in App. A of Ref.~\cite{SP},
we find that there are 8 even-even $U(1)_A$ singlets:
\begin{eqnarray}
\left\{[S\times S]-[P\times P]\right\} &=& \half(\CF_1+\CF_2)\,,\\
\left\{[V_\mu\times V_\mu]-[A_\mu\times A_\mu]\right\} 
&=& \half(\CF_1-\CF_2)\,,\\
\left\{[S\times P]-[P\times S]\right\}\,,\\
\left\{[V_\mu\times A_\mu]-[A_\mu\times V_\mu]\right\}\,,\\
\left\{[V\times V]-[A\times A]\right\}\,,\\
\left\{[V\times A]-[A\times V]\right\}\,,\\
\left\{[S\times T]-[P\times T]\right\}\,,\\
\left\{[T\times S]-[T\times P]\right\}\,.
\end{eqnarray}
The first two are related to two of Luo's operators, while the
remaining six are new.
Note that no $[T\times T]$ operators are $U(1)_A$ singlets.

The continuum versions of these operators (which appear in the effective
Lagrangian for unimproved or partially improved staggered fermions)
can be obtained simply by replacing the $\chi$'s with $Q$'s, and
removing the bar on the matrices. In the text we use the same notation
for the corresponding continuum operators, e.g. $[S\times V]$, as we
do for the lattice operators.

\section{Mapping of operators in $S_6^{\rm FF(A)}$ into $\CV_\chi^6$}
\label{app:matching}

The operators in $S_6^{\rm FF(A)}$ break into three 
classes according to the spin of the bilinears.

\subsection{Operators with spin structure $V$ or $A$}

The general form of these operators is a linear
combination of $[V\times F]$ and $[A\times F]$,
with $F$ denoting the flavor.
The flavor structure of these operators is a sum of terms of the form
\begin{equation}
\CO_F= \pm \sum_\mu \left(\bar {Q_R} (\gamma_\mu\otimes F_R) Q_R \pm 
\bar{Q_L} (\gamma_\mu\otimes F_L) Q_L \right)^2
\label{eq:OFform}
\end{equation}
where $F^{L,R}$ are hermitian matrices in flavor space.
Both $\pm$ signs are positive (resp. negative) if the spin is 
$V$ (resp. $A$). For the axial operators,
the internal minus sign is due to the extra $\gamma_5$,
while the external sign appears because for spin $A$ the Dirac matrix is
$i \gamma_\mu\gamma_5$ so as to be hermitian.
To determine the form of the corresponding operator in $\CL_\chi^6$,
we promote $F_{L,R}$ to spurion fields.
If they transform as $F_L\to L F_L L^\dagger$
and $F_R \to R F_R R^\dagger$, then $\CO_F$ is invariant under
chiral transformations.
We then build all the operators out of $\Sigma$ which are chiral singlets,
quadratic in $F$ (linear and cubic terms are forbidden by 
the $F\to -F$ symmetry, while quartic terms would be of $O(a^4)$),
and parity invariant ($F_L\leftrightarrow F_R$ 
and $\Sigma\leftrightarrow \Sigma^\dagger$). 
At the end, we set $F_L=F_R=F$, where $F$ is the flavor matrix
appearing in the operator under consideration
(e.g. $\xi_5$ for $[V\times P]$).

It turns out that there is only one
non-trivial operator into which $\CO_F$ can map
\begin{equation}
\CO_F \rightarrow c \Tr(F_L\Sigma F_R\Sigma^\dagger) \,.
\end{equation}
The constant, $c$, is unknown, but has the same magnitude
(and sign) for both $V$ and $A$ spins.
The sign is the same because the overall sign in (\ref{eq:OFform})
cancels with the ``internal'' sign 
coming from the fact that this operator contains
one factor each of $F_L$ and $F_R$.
A second operator allowed by the symmetries,
\begin{equation}
 \Tr(F_L)\Tr(\Sigma F_R \Sigma^\dagger) +
\Tr(F_R) \Tr(\Sigma^\dagger F_L \Sigma)  \,,
\end{equation}
turns out to be
a field independent constant when one sets $F_L=F_R=F$.
This is either because $\Tr(F)=0$ (true for flavor $P$ and $T$)
or because $\Sigma \Sigma^\dagger=1$ (for flavor singlet bilinears).
Finally, the operators $\Tr(F_L^2)\Tr(\Sigma\Sigma^\dagger)$ and
$\Tr(F_L^2\Sigma\Sigma^\dagger)$ are trivial for any $F$.

Thus we find that the operators map into the following terms 
in the potential $\CV_\chi^6$:
\begin{eqnarray}
\,[V\times S]\,\ {\rm and}\ \,[A\times S] &\longrightarrow& 
	4 c\,,\vphantom{\Tr} \label{eq:VSmap}\\
\,[V\times P]\,\ {\rm and}\ \,[A\times P] &\longrightarrow& 
	c \Tr(\xi_5\Sigma \xi_5 \Sigma^\dagger)\,, \label{eq:VPmap}\\
\,[V\times T]\,\ {\rm and}\ \,[A\times T] &\longrightarrow&
c \sum_{\mu<\nu} 
\Tr(\xi_{\mu\nu}\Sigma \xi_{\nu\mu} \Sigma^\dagger) \,,\label{eq:VTmap}\\
\lefteqn{
([V\times V]-[A\times A]) \ {\rm and}\ 
([A\times V]-[V\times A])} \hspace{2in} \nonumber \\ 
&\longrightarrow&
c \sum_{\nu} \left\{
\Tr(\xi_\nu\Sigma \xi_\nu \Sigma^\dagger) 
- \Tr(\xi_{\nu5}\Sigma \xi_{5\nu} \Sigma^\dagger) 
\right\}
\,.\label{eq:V-Amap}
\end{eqnarray}
The fact that the same constant $c$ appears in all these 
transcriptions is crucial for the last of these mappings,
eq.~(\ref{eq:V-Amap}).
A check on this result
is that the operator is invariant under the axial $U(1)$ 
only if the relative coefficient of the two terms is $-1$.
Apart from this, the fact that the same constant $c$ appears in
all transcriptions does not lead to useful relations because the
coefficients of the underlying quark operators are different and
unknown.

\subsection{Operators with spin structure $S$ or $P$}

The chiral structure of operators of the form $[S\times F]$ 
and $[P\times F]$ is
\begin{equation}
\CO'_F= \left(\bar {Q_L} (1\otimes \tilde F_L) Q_R \pm 
\bar{Q_R} (1\otimes \tilde F_R) Q_L \right)^2 \,,
\end{equation}
with the $\pm$ sign corresponding to $S$ or $P$.
Here the spurions must transform as
$\tilde F_L \to L \tilde F_L R^\dagger$ and
$\tilde F_R \to R \tilde F_R L^\dagger$.
The non-trivial operators into which $\CO'$ maps are
\begin{eqnarray}
\CO'_F&\longrightarrow&
\pm c_1 \Tr(\tilde F_R \Sigma) \Tr(\tilde F_L \Sigma^\dagger) 
\nonumber\\
&&\mbox{} + c_2
\left[\Tr(\tilde F_R \Sigma)^2 + \Tr(\tilde F_L \Sigma^\dagger)^2 \right] 
\nonumber\\
&&\mbox{}
+ c_3 \left[\Tr(\tilde F_R \Sigma \tilde F_R \Sigma) + 
\Tr(\tilde F_L \Sigma^\dagger \tilde F_L \Sigma^\dagger) \right] \,.
\label{eq:oprimemap}
\end{eqnarray}
From this we find that
\begin{eqnarray}
\,[S\times V]\ \ {\rm and}\ \ [P\times V] &\longrightarrow& 
\pm c_1 \Tr(\xi_\nu \Sigma)\Tr(\xi_\nu\Sigma^\dagger)
\nonumber \\
&& \mbox{} + c_2 \left[\Tr(\xi_\nu \Sigma)^2 + h.c. \right]  \nonumber\\
&& \mbox{} + c_3 \left[\Tr(\xi_\nu\Sigma\xi_\nu \Sigma) + h.c.\right] 
\label{eq:SVmap}\\
\,[S\times A]\ \ {\rm and}\ \ [P\times A] &\longrightarrow& 
 \pm c_1 \sum_\nu \Tr(\xi_{\nu5} \Sigma)\Tr(\xi_{5\nu}\Sigma^\dagger)
\nonumber \\ && \mbox{} 
+ c_2 \sum_\nu 
\left[\Tr(\xi_{\nu5} \Sigma)\Tr(\xi_{5\nu}\Sigma) + h.c. \right] 
\nonumber \\ && \mbox{} 
+ c_3 \sum_\nu 
\left[\Tr(\xi_{\nu5}\Sigma\xi_{5\nu} \Sigma) + h.c.\right] 
\,. \label{eq:SAmap}
\end{eqnarray}
For the operators which are linear combinations, we find
\begin{eqnarray}
([S\times S]-[P\times P]) &\longrightarrow&
c_1 \left[ \Tr(\Sigma)\Tr(\Sigma^\dagger)+ 
\Tr(\xi_5\Sigma)\Tr(\xi_5\Sigma^\dagger) \right]
\nonumber \\ && \mbox{}
+ c_2 \left[ \Tr(\Sigma)^2 -\Tr(\xi_5\Sigma)^2 + h.c.\right]
\nonumber \\ && \mbox{}
+ c_3 \left[\Tr(\Sigma^2) -\Tr(\xi_5\Sigma\xi_5\Sigma) + h.c.\right] \,,
\label{eq:SSmap}
\end{eqnarray}
with $([S\times P]-[P\times S])$ mapping into the same three
operators except that the $c_2$ and $c_3$ terms change sign.
A check on these results is that the
relative plus sign in the $c_1$ term and minus signs in
the other terms are those needed to make these operators 
invariant under $U(1)_A$.
Finally, for the flavor $T$ only the $c_1$ term survives
\begin{equation}
([S\times T]-[P\times T]) \longrightarrow
2 c_1 \sum_{\mu<\nu} \Tr(\xi_{\mu\nu}\Sigma)\Tr(\xi_{\nu\mu}\Sigma^\dagger)
\,.
\label{eq:STmap}
\end{equation}

\subsection{Operators with spin structure $T$}

There are two such operators,
$[T\times V]$ and $[T\times A]$.
These have a slightly different chiral structure 
to that of the $S$ and $P$ operators,
\begin{equation}
\CO''_F= \sum_{\mu<\nu}
\left(\bar {Q_L} (\gamma_{\mu\nu}\otimes \tilde F_L) Q_R\right)^2 +
\left(\bar{Q_R} (\gamma_{\nu\mu}\otimes \tilde F_R) Q_L \right)^2 \,.
\end{equation}
In particular, there are no cross terms between $\bar LR$ and 
$\bar RL$ bilinears.
It follows that the mapping is as in eq.~(\ref{eq:oprimemap})
except that there is no $c_1$ term, and the other terms have new
coefficients $c'_2$ and $c'_3$.
From this we find that
\begin{eqnarray}
\,[T\times V] &\longrightarrow &
+ c'_2 \sum_\nu \left[\Tr(\xi_\nu \Sigma)^2 + h.c. \right] 
\nonumber \\ && \mbox{}
+ c'_3 \sum_\nu \left[\Tr(\xi_\nu\Sigma\xi_\nu \Sigma) + h.c.\right] 
\,,
\label{eq:TVmap}\\
\,[T\times A] &\longrightarrow &
- c'_2 \sum_\nu \left[\Tr(\xi_{\nu5} \Sigma)^2 + h.c. \right] 
\nonumber \\ && \mbox{}
+ c'_3 \sum_\nu \left[\Tr(\xi_{\nu5}\Sigma\xi_{5\nu} \Sigma) + h.c.\right] 
\,.
\label{eq:TAmap} 
\end{eqnarray}

\end{document}